\documentclass[12pt,twoside,fleqn]{article}
\usepackage{epsfig}
\usepackage{psfig}
\def\lsim{\raise0.3ex\hbox{$<$\kern-0.75em\raise-1.1ex\hbox{$\sim$}}}
\def\gsim{\raise0.3ex\hbox{$>$\kern-0.75em\raise-1.1ex\hbox{$\sim$}}}
\makeatletter
\@addtoreset{equation}{section}
\makeatother

\newcommand{\beqn} {\begin{equation}}
\newcommand{\eqn} {\end{equation}}

\newcommand{\slsh}[1] {#1\kern-.43em/}
\newcommand{\real}{{\sf I}\kern-.12em{\sf R}}
\newcommand{\comp}{{\sf I}\kern-.48em{\sf C}}
\newcommand{\nin} {\in\kern-.6em/}
\newcommand{\Tr} {\mbox{Tr}}

\def\MEF{m_{\rm eff}}\def\mef{\ifmmode\MEF\else$\MEF$\fi}

\def\LAT{{\NP B (Proc.\ Suppl.)\ }}
\def\NP{{Nucl.\ Phys.\ }}

\def\PL{{Phys.\ Lett.\ }}
\def\PR{{Phys.\ Rev.\ }}

\begin{document}
\thispagestyle{empty}
%
 \mbox{} \hfill BI-TP 96/10\\
 \mbox{} \hfill WUB 96-5\\
 \mbox{} \hfill February 1996\\
\begin{center}
\vspace*{1.0cm}
{{\large \bf Gauge Boson Masses in the 3-d, SU(2) Gauge-Higgs
Model }
 } \\
\vspace*{1.0cm}
{\large F. Karsch$^1$, T. Neuhaus$^2$, A. Patk\'os$^3$
and J. Rank$^1$}
\\
\vspace*{1.0cm}
{\normalsize
$\mbox{}$ {$^1$ Fakult\"at f\"ur Physik, Universit\"at Bielefeld, P.O. Box 100131,
D-33501 Bielefeld, Germany  \\
$^2$ Fachbereich Physik, Gesamthochschule Universit\"at Wuppertal,
D-42097 Wuppertal, Germany  \\
$^3$ Institute of Physics, E\"otv\"os University, Budapest, Hungary
}}\\
\vspace*{1.8cm}
{\large \bf Abstract}
\end{center}
\setlength{\baselineskip}{1.3\baselineskip}

We study gauge boson propagators in the symmetric and symmetry broken
phases of the 3-d, $SU(2)$ gauge-Higgs model. Correlation functions
for the gauge fields are calculated in Landau gauge. They are found
to decay exponentially at large distances leading to a non-vanishing
mass for the gauge bosons.
We find that the W-boson screening mass drops in the symmetry broken phase
when approaching the critical temperature.
In the symmetric phase the screening mass stays small and is
independent of the scalar--gauge coupling (the hopping parameter).
Numerical results coincide with corresponding calculations performed for
the pure gauge theory. We find $m_w = 0.35(1)g^2T $ in this phase
which is consistent with analytic calculations based on gap equations.
This is, however, significantly
smaller than masses extracted from gauge invariant vector boson correlation
functions. As internal consistency check we also have calculated
correlation functions for gauge invariant operators leading to scalar
and vector boson masses. Finite lattice size effects
have been systematically analyzed on lattices of size $L^2\times L_z$ with
$L=4-24$ and $L_z = 16 - 128$.

\newpage
\setcounter{page}{1}
\section{Introduction}

The standard model of electroweak interactions predicts the existence of a
phase transition to a high temperature symmetric phase \cite{Lin72}.
The knowledge of fluctuation spectra in the high temperature phase is
essential for understanding its physics.

In the high temperature phase the thermal contribution to the
screening masses of $W$-gauge bosons dominates the contribution (if any)
of the vacuum expectation value of the Higgs field. One has to
distinguish here between the screening scales resulting from
electric and magnetic gauge field fluctuations.
The leading order electric screening mass is ${\cal O}(gT)$. It is
essentially a perturbative quantity, determined by the internal consistency
of the resummed perturbative treatment of the thermodynamics of the system
\cite{Arn93,Karun}.
Beyond leading order it requires a careful non-perturbative definition
\cite{Arn96}.

The non-vanishing magnetic screening mass does play an important role in
controlling the infrared behaviour of the electroweak theory at high
temperature and does influence the nature of the electroweak phase transition
itself. For instance, it is expected that the existence or non-existence of a
first order phase transition in the electroweak theory crucially
depends on the magnitude of the thermal magnetic mass of the $W$-boson
\cite{Buc94,Esp93}. The role of this mass in the symmetric high
temperature phase of the electroweak theory is similar to that of the
magnetic mass generated for the gluons in the high temperature, deconfined
phase of QCD. Also, the thermal magnetic mass is crucial for the infrared
behaviour of QCD at high temperature. In both cases these masses are
expected to be of ${\cal O} (g^2 T)$ \cite{Lin80} up to possible logarithmic
corrections \cite{Kal92}.

The magnetic masses for the gauge bosons in the $SU(2)$ gauge-Higgs
model as well as in QCD are not calculable within the context of high
temperature perturbation theory.
For instance, one might attempt to apply some sort of resummation to the
magnetic sector, represented by a 3-dimensional effective theory in
both cases \cite{Bra95}. A perturbative calculation of the free energy with a magnetic
mass of ${\cal O}(g^2T)$ introduced by adding and subtracting
a corresponding term to the Lagrangian would lead to a series in
$g^2T/m_w(T)$. All evidence gathered so far on this ratio suggest
a large value for it. Any perturbative determination of it is therefore
expected to fail.

Some form of a non-perturbative
approach is needed. In the case of the $SU(2)$ gauge-Higgs model
there have been various Monte Carlo calculations in which the thermal
vector boson as well as the Higgs masses have been calculated with help of gauge
invariant correlators of appropriate quantum numbers
\cite{Fod95,Ilg95,Kaj95,Phi96}.

Another attempt is based on an analytic treatment
of coupled gap equations for the scalar and vector propagators on the
mass shell \cite{Buc95,PhiSint}. Such an analysis leads to the conjecture
that also in the high temperature phase the magnetic
$W$-boson mass is generated essentially by a Higgs-type phenomenon. The
difference being that  the expectation value of the order
parameter is much smaller. It therefore is intuitively very appealing to
continue the use of the same gauge invariant operators for the calculation of
the magnetic mass  (and also of the Higgs-mass), like in the low temperature
phase.

The numerical calculations of gauge invariant correlators, however, do lead to
a thermal mass for the vector boson which is substantially larger than
the result of the analysis of gap equations in Landau gauge.
This situation is somewhat similar to the case
of $SU(N)$ gauge theories where the analysis of gauge invariant {\it
glueball} operators with the quantum numbers of the {\it gluon} \cite{Gro94}
does lead to much larger screening masses than the direct calculation of the
gluon propagator in Landau gauge \cite{Hel95}. In this case the observed
discrepancy was, however, expected. The
gauge invariant correlation functions correspond to glueball states
at low temperature and {\it melt} into several decoupled gluons with an
effective thermal mass in the high temperature deconfined phase. The gauge
invariant glueball correlation functions thus describe a multiple gluon state
in the high temperature phase. The elementary thermal mass is only visible
in a direct calculation of the gluon propagator.

In the $SU(2)$ gauge-Higgs model, however, on the basis of the above argument
the discrepancy between analytic results
and numerical calculations in the symmetric phase is somewhat
unexpected.
An explanation could be that analytic
calculations are less stable in the symmetric phase.
However, this discrepancy may also hint at a situation
similar to the case of QCD, i.e. the gauge invariant operators may not
project onto single $W$-boson states in the high temperature symmetric phase.
Also the agreement of the bound state model of Dosch {\it et al.}
with the screening masses obtained from the spectroscopy of gauge invariant
operators \cite{Dos95} points to such an interpretation.

In order to get closer to a clarification of this problem we
need a more detailed quantitative understanding of both the behaviour of
correlation functions for gauge fields in fixed gauges as well as that
of gauge independent correlation functions with quantum numbers of the
gauge bosons.

It is the purpose of this paper to study in detail the behaviour of
correlation functions for the gauge potentials in Landau gauge.
This will be done within the context of the 3-dimensional
$SU(2)$ gauge-Higgs model which is obtained as an effective theory for the
finite temperature electroweak model by integrating out heavy static modes
corresponding to the zeroth component of the gauge fields \cite{Kaj96}.
We extract from the exponential fall-off of correlation functions of the
gauge fields the $W$-boson magnetic mass in Landau gauge and compare with
corresponding calculations of gauge invariant operators with quantum numbers
of the $W$-boson as well as the Higgs boson. We have performed these
calculations on a large number of different lattice sizes in order to
control finite size effects. In addition we have performed calculations
for the pure $SU(2)$ gauge theory in order to test the conjecture that the
$W$-boson mass in the symmetric phase is closely related to the magnetic
mass in the $SU(2)$ gauge theory. We do not address the problem of gauge
invariance of the masses extracted from the gauge boson propagators and
the related issue of influence of Gribov copies in this paper.
These problems have been discussed in the context of calculations for
photon and gluon propagators \cite{Nak91}.

The paper is organized as follows. In section 2 we give the basic
definitions for the 3-d gauge-Higgs model and its relation to the
(3+1)-dimensional finite temperature $SU(2)$ Higgs model. In section 3 we
discuss the calculation of the $W$-boson propagator in Landau gauge.
The analysis of gauge invariant scalar and vector correlation functions
is presented in section 4. Finally we give our conclusions in section 5.
\section{Reduced EW-model}

The physics of the longest range fluctuations of the finite temperature
electroweak theory is described by an effective  3-d theory which is matched
to the complete model at the distance scale $a\sim (gT)^{-1}$
\cite{Kaj96,Kaj93,Kaj94,Jak94}.
Though this effective theory coincides formally with the superrenormalisable
3-d gauge--Higgs model, the optimal quantitative description of the original
physics is expected to be obtained by choosing the lattice spacing according to
$\Theta\equiv aT\sim 1$.

The lattice formulation of the effective theory is given by
\begin{eqnarray}
&
S^{3D}_{lat}={\beta \over 2}\sum_P{\rm Tr}U_P(x)+{1\over 2} \sum_{x,i}
{\rm Tr}\Phi^\dagger_xU_{x,i}\Phi_{x+\hat{i}}-{1\over 2\kappa}\sum_x{1\over 2}
{\rm Tr} \Phi^\dagger_x\Phi_x\nonumber\\
&
-{\lambda_3\over 24}\sum_x({1\over 2}{\rm Tr}
\Phi^\dagger_x\Phi_x)^2,
\end{eqnarray}
where $U_P$ is the standard plaquette variable of the $SU(2)$ lattice gauge
theory and $\Phi_x$ is a complex $2\times 2$ matrix which in terms of
the real weak isosinglet-triplet decomposition of the
complex Higgs doublet is given by $\Phi_x \equiv \Phi_0+\tau_i\Phi_i$.

The relationship of the dimensionless lattice couplings $\beta,~\lambda_3,~
\kappa$ to the couplings of the original $T=0$, $SU(2)$ gauge--Higgs system
is given by the following sequence of equations:
\begin{equation}
\beta ={4\over g_3^2\Theta},~~~ g_3^2=g^2(1-{g\over 20\pi}\sqrt{5\over 6}),
\label{beta}
\end{equation}
\begin{equation}
\lambda_3=\biggl(
{3\over 4}{m_H^2\over m_W^2}g^2-{27\over 160\pi}\sqrt{5\over 6}g^3
\biggr)\Theta,~~m_H^2={\lambda\over 3}v^2,~~m_W^2={g^2\over 4}v^2,
\label{lambda}
\end{equation}
\begin{equation}
{1\over \kappa}=m^2a^2+({3\over 16}g^2+{1\over 12}
\lambda -{3g^3\over 16\pi}\sqrt{5\over 6})\Theta^2
-\Theta\Sigma (L)({3\over 2}g^2+\lambda -{15g^3\over 32\pi}\sqrt
{5\over 6})+6.
\label{kappa}
\end{equation}
In these equations $g,\lambda, m^2$ represent the renormalised parameters
of the original theory and $m_H,~m_W$ are the $T=0$ masses of the Higgs and
$W$ bosons. $\Sigma (L)$ is a slightly size dependent geometrical
factor which is  known exactly for any lattice size $L$. In particular one
obtains in the infinite volume limit
$\Sigma (\infty )= 0.252731$.

The above relations are obtained when one integrates first over the non-static
Matsubara modes with 1-loop accuracy and in a second reduction step also
over the static $A_0$ mode \cite{Kaj94}.
Then the coupling relations are accurate to ${\cal O}(g^3,\lambda^{3/2})$.
The $A_0$-integration has been realised with help of an
iterated 1-loop calculation \cite{Pol95}. Although it will not be relevant
for the following discussion it is worth to note that
the ${\cal O}(g^3\Theta)$ term in Eq.~(\ref{kappa}) which has been
estimated with this iterative technique deviates from the result given
in \cite{Kaj94}.

Our numerical calculations have been performed with the parameters
\begin{equation}
\beta =9.0 \quad , \quad \lambda_3 = 0.313646~~,
\end{equation}
at the physical $W$-boson mass, $m_W=80.6$ GeV.
Choosing also $\Theta =1$
this does correspond to a Higgs mass of about 80~GeV (Eq.~(\ref{lambda})).
Eq.~(\ref{kappa}) does then relate
$\kappa$ to the ratio $m^2/T^2$. After finding $\kappa_c$,
its value is easily translated into $T_c$ (with one more
input of $v=246$ GeV). This can be performed for each
value of $\kappa$. Our correlation measurements covered the range
$0.170\leq \kappa\leq 0.180$. The $\kappa-$range can be translated with
help of Eq.~(\ref{kappa}) into a temperature interval around the
critical value, $0.6~T_c~\lsim~T~\lsim~(4-5)~T_c$.
We expect that more accurate
mappings between the couplings will not modify qualitatively the temperature
interval covered.

We have chosen to work at a rather large value of the Higgs mass.
For this choice of parameters the nature of the phase transition has not yet
been clarified. It could be either
continuous or only very weakly first order. The study of the gap equations
in \cite{Buc95} indicates a smooth crossover. We shall give an
argument below in favor of a second order transition. One therefore should be 
able to find a critical value of the hopping
parameter at which the symmetry restoring takes place. This has been
determined by us as
\begin{equation}
\kappa_c = 0.17467(2)~~.
\end{equation}
We will report in more detail about the determination of $\kappa_c$ and
an analysis of the order of the phase transition elsewhere.

\section{W-boson propagator in Landau gauge}

\subsection{Landau gauge fixing}

Our analysis of the $W$-boson propagator in Landau gauge closely follows the
approach used in the calculation of the gluon propagator in finite
temperature QCD \cite{Hel95,Kar95}. We define the gauge fields,
$A_{\mu}({\bf x})$, from the $SU(2)$ link variables, $U_{x,\mu}$
\begin{equation}
A_{\mu}({\bf x}) = {i \over 2g}
\left( U^{\dagger}_{x,\mu} - U_{x,\mu} \right) \ .
\label{a_mu}
\end{equation}
The Landau gauge condition $|\partial_{\mu}A^{\mu}({\bf x})|^2 = 0$ is
then realized on each lattice configuration by maximizing
the trace of the link fields, $U_{x,\mu}$,
\begin{equation}
\Sigma = \sum_{x, \mu} \Tr \, \left[ U_{x,\mu} + U^{\dagger}_{x,\mu}
\right]~~.
\label{spur_summe_1b}
\end{equation}
The maximization has been performed using an overrelaxation algorithm
combined with a FFT-algorithm \cite{Dav88,Man88} until the Landau gauge
condition has been satisfied within an accuracy of $10^{-9}$. Typically
this required about 500 iterative maximization steps.

On the gauge fixed configurations we analyze correlation functions of gauge
fields averaged over $(x,y)$-planes,
\begin{equation}
\tilde{A}_\mu (z) = \sum_{x,y} A_\mu (x,y,z)~~,
\label{Aave}
\end{equation}
in order to improve the projection onto the zero momentum excitations.
The correlation functions of these averaged fields are then calculated in
the transverse $z$-direction,
\begin{equation}
G_w (z) = \left\langle \Tr \; \tilde{A}_\mu (0)\tilde{A}_\mu (z)
\right\rangle~~.
\label{corr}
\end{equation}
For large separations $z$ these correlation functions do project onto the
$W$-boson propagator mass.

\subsection{W-boson propagator}

We have analyzed the $W$-boson propagator in Landau gauge in the 3-d $SU(2)$
gauge-Higgs model.
Calculations have been performed for a large number of hopping parameter values.
We have used an overrelaxed heat-bath algorithm and performed typically
50.000 iterations\footnote{We call an iteration a
combination of 4 overrelaxation steps followed by one heat-bath update.
The heat-bath update of the scalar fields was optimized by shifting a 
suitably chosen quadratic term of the Higgs field from the accept reject 
decision into the generation of the cartesian gaussian components.}
per $\kappa$-value.
After every tenth iteration we have then fixed the Landau gauge and calculated
the gauge fixed correlators as well as a set of gauge invariant operators.
Most calculations have been performed on a lattice of size
$16^2\times 32$. In order to get control over finite lattice size effects
we have
performed additional calculations on lattices of size $L^2 \times 32$
with $L$ ranging from 4 to 24 as well as $16^2\times L_z$ with $L_z$, ranging
from 16 to 128. These calculations have been performed at two values of
the hopping parameter below and above $\kappa_c$.
The statistics accumulated for our detailed finite size analysis at these
$\kappa$-values is summarized in Table~\ref{corrfit.tab}.

The analysis of the volume dependence
allowed us to select a suitable ansatz for the fits of correlation
functions on the $16^2\times 32$ lattice which minimize finite size
effects in the determination of the $W$-boson propagator mass.
\begin{figure}[htb]
  \epsfig{file=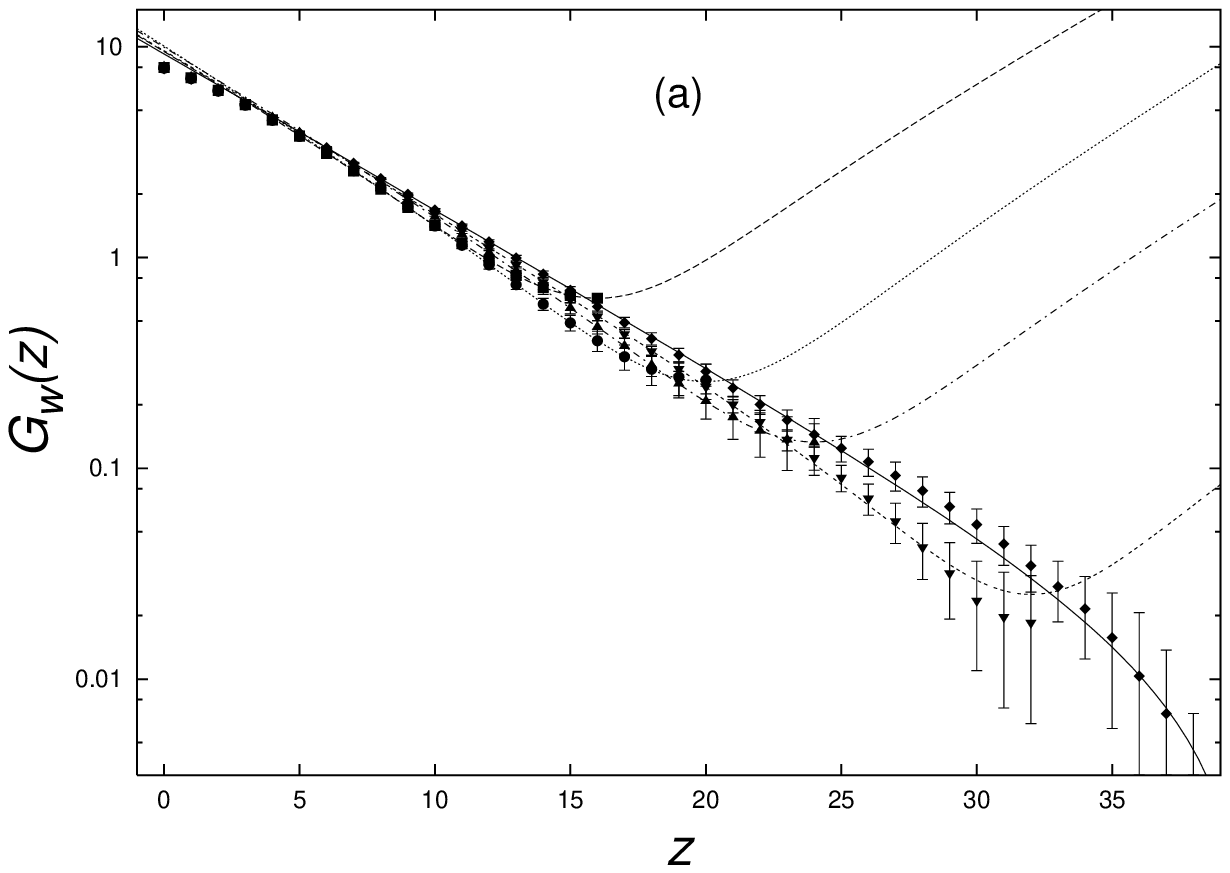, width=73mm, height=65mm}
  \hfill
  \epsfig{file=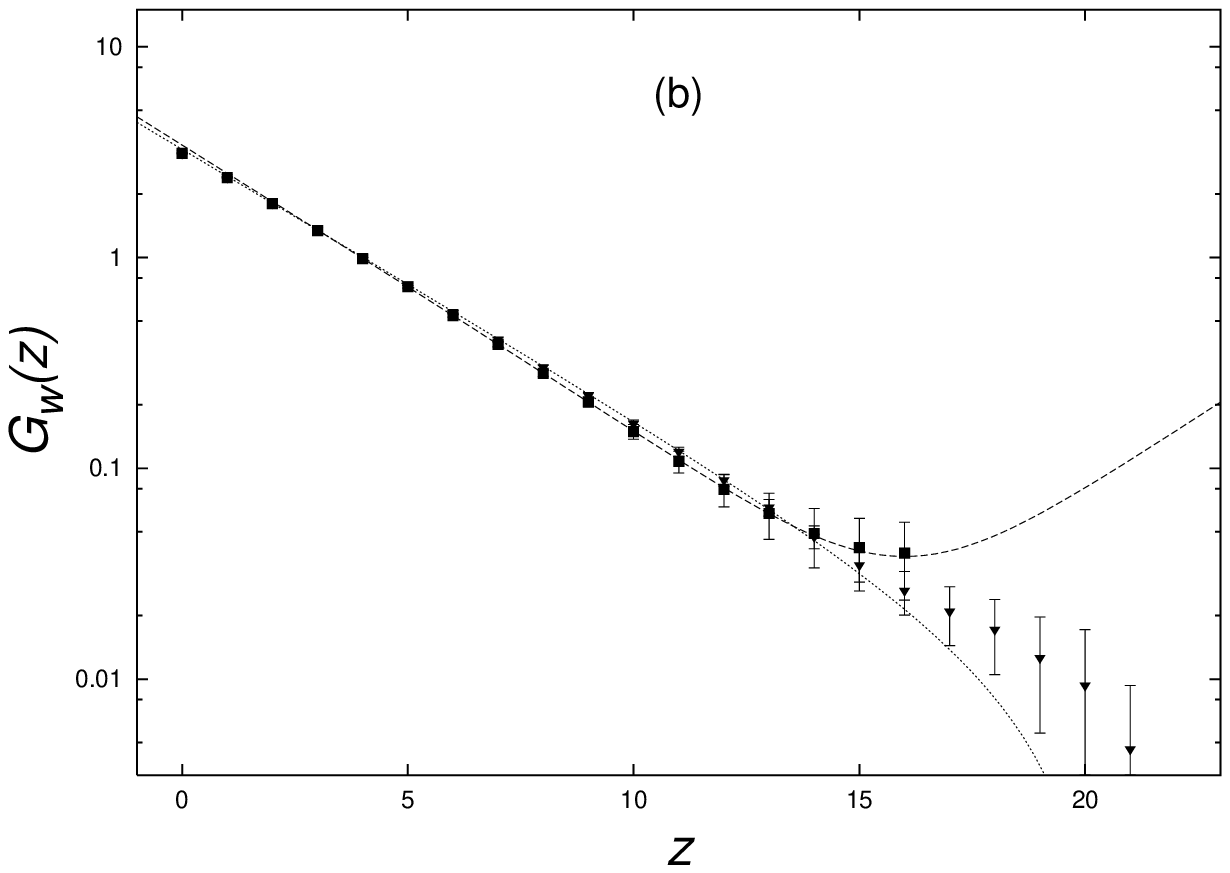, width=73mm, height=65mm}
\caption{Gauge field correlation functions on $16^2 \times L_z$ lattices
with $L_z = 32$ (squares), 40 (circles), 48 (upper triangles), 64 (lower
triangles) and 128 (diamonds). Shown are correlation functions in the
symmetric phase at $\kappa=0.1745$ (a) and the symmetry broken phase at
$\kappa=0.17484$ (b). The curves give fits for $z\ge 8$. The fitting
parameters are listed in Table 1. }
\label{wcorr.fig}
\end{figure}
In Fig.~\ref{wcorr.fig} we show the correlation function $G_w (z)$ defined in
Eq.(\ref{corr}) for various lattice sizes at $\kappa = 0.1745$ and $\kappa =
0.17484$.
The correlation functions show a slower decay at short
distances which also has been observed in the analysis of the gluon
propagator \cite{Hel95,Nak95}. This behaviour is also evident from the
analysis of local masses which approach a plateau at large distances from
below.
We define local masses in two different ways either as solution of the equation

\beqn
{ G_w (z-1) - G_w (z) \over G_w (z) -G_w (z+1) } =
{ G_w^{fit} (z-1) - G_w^{fit} (z) \over G_w^{fit} (z) - G_w^{fit} (z+1) }\quad ,
\label{wlocala}
\eqn
or
\beqn
{ G_w (z-1) \over G_w (z)} = { G_w^{fit} (z-1) \over G_w^{fit} (z) }\quad .
\label{wlocalb}
\eqn
Here $G_w (z)$ denotes the calculated values for the correlation functions
and
\beqn
G_w^{fit} (z) = A~\biggr(\exp{(-m_w z)} + \exp{(-m_w (L_z - z))}\biggl)~+~B~~.
\label{wfit}
\eqn
\begin{table}
\begin{center}
\begin{tabular}{|r|l|l|l|r|}
\hline
$L_z$ & \multicolumn{1}{|c|}{$m_w$} & \multicolumn{1}{|c|}{$A$} &
\multicolumn{1}{|c|}{$B$} & \# iterations \\
\hline
\multicolumn{5}{|c|}{$\kappa = 0.1745$} \\
\hline
   32    &     0.166(~7)     &  10.3~(2) & -0.80(15)  & 190.000 \\
   40    &     0.194(14)     &  10.9~(6) & -0.18(13)  & 40.000 \\
   48    &     0.179(11)     &  10.1~(9) & -0.14(~8)  & 40.000 \\
   64    &     0.174(~9)     &   9.3~(8) & -0.045(22) & 90.000  \\
  128    &     0.163(12)     &   8.6~(9) & -0.012(13) & 60.000  \\
\hline
\multicolumn{5}{|c|}{$\kappa = 0.17484$} \\
\hline
   32    &     0.308(~6)     &   3.4~(1) & -0.012(19) & 80.000  \\
   64    &     0.291(11)     &   3.2~(2) & -0.009(~4) & 40.000  \\
\hline
\end{tabular}
\end{center}
\caption{Results of fits to the correlation functions shown in
Fig.~1. }
\label{corrfit.tab}
\end{table}

\begin{figure}[htb]
\begin{center}
  \epsfig{file=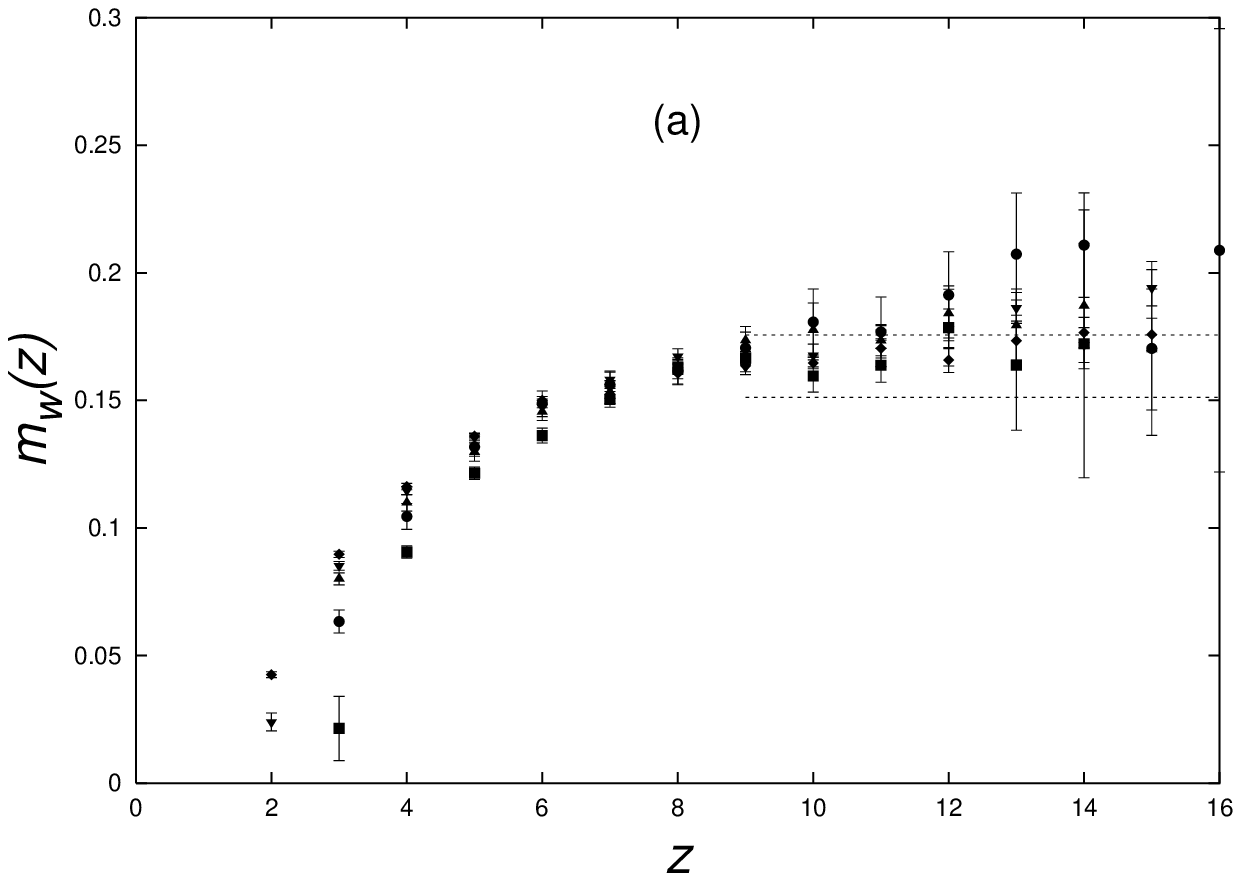, width=73mm, height=65mm}
  \hfill
  \epsfig{file=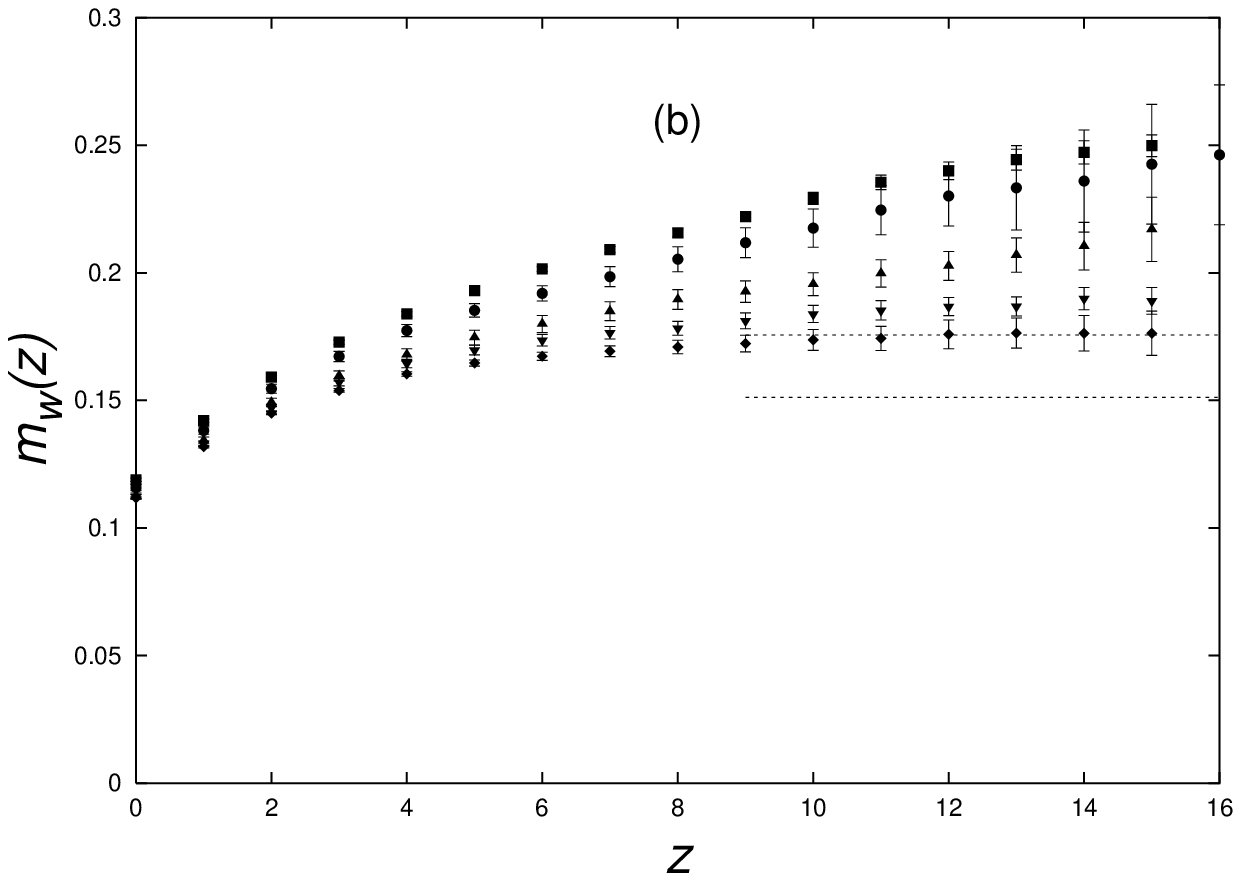, width=73mm, height=65mm}
\end{center}
\caption{Local masses calculated at $\kappa = 0.1745$
from the correlation functions shown in Fig.~1a. In (a) we show local masses
extracted according to Eq.~(3.5) while (b) gives the result according to
Eq.(3.6). The latter does assume $B=0$. The horizontal lines give the error band
resulting from the fit on a $16^2\times 128$ lattice. }
\label{wlocal.fig}
\end{figure}

While the ansatz given in Eq.~(\ref{wlocala}) is independent of the constant
$B$ the second version given in Eq.~(\ref{wlocalb}) is not. In the latter
case we use $B \equiv 0$ to define the local masses.
Results for these are shown in Fig.~\ref{wlocal.fig}. We note
that there is no apparent dependence on $L_z$ visible in the analysis based
on Eq.~(\ref{wlocala}) while there is a significant
volume dependence if we use ths second ansatz. We do, however, find
that both forms yield consistent results for $L_z \ge 64$. This suggests
that we can minimize finite lattice size effects by allowing for a constant
in our ansatz for a global fit to the correlation functions. We also see
from Fig.~\ref{wlocal.fig} that the local masses do
develop a plateau for $z \gsim 8$.
We therefore have fitted the correlation functions only for distances $z \ge 8$
using the ansatz given in Eq.~(\ref{wfit}).
The fit results for the correlation functions shown in
Fig.~\ref{wcorr.fig} are summarized in Table~\ref{corrfit.tab}. We note that
these fits yield values for the propagator masses $m_w$ which are within
errors independent of $L_z$. Moreover, the constant $B$ rapidly drops to
zero with increasing $L_z$. We find that it is well described by an
exponential decrease, $B \sim \exp{(-0.1 L_z)}$. We also note that the
constant $B$ is consistent with being zero in the symmetry broken phase
already for $L_z = 32$.
A similar analysis has been performed for the dependence of the correlation
functions on the transverse lattice size. In that case simulations have been 
performed on lattices of size $L^2 \times 32$ with $L$ ranging from 4 to 24.

\begin{figure}[htb]
\begin{center}
  \epsfig{file=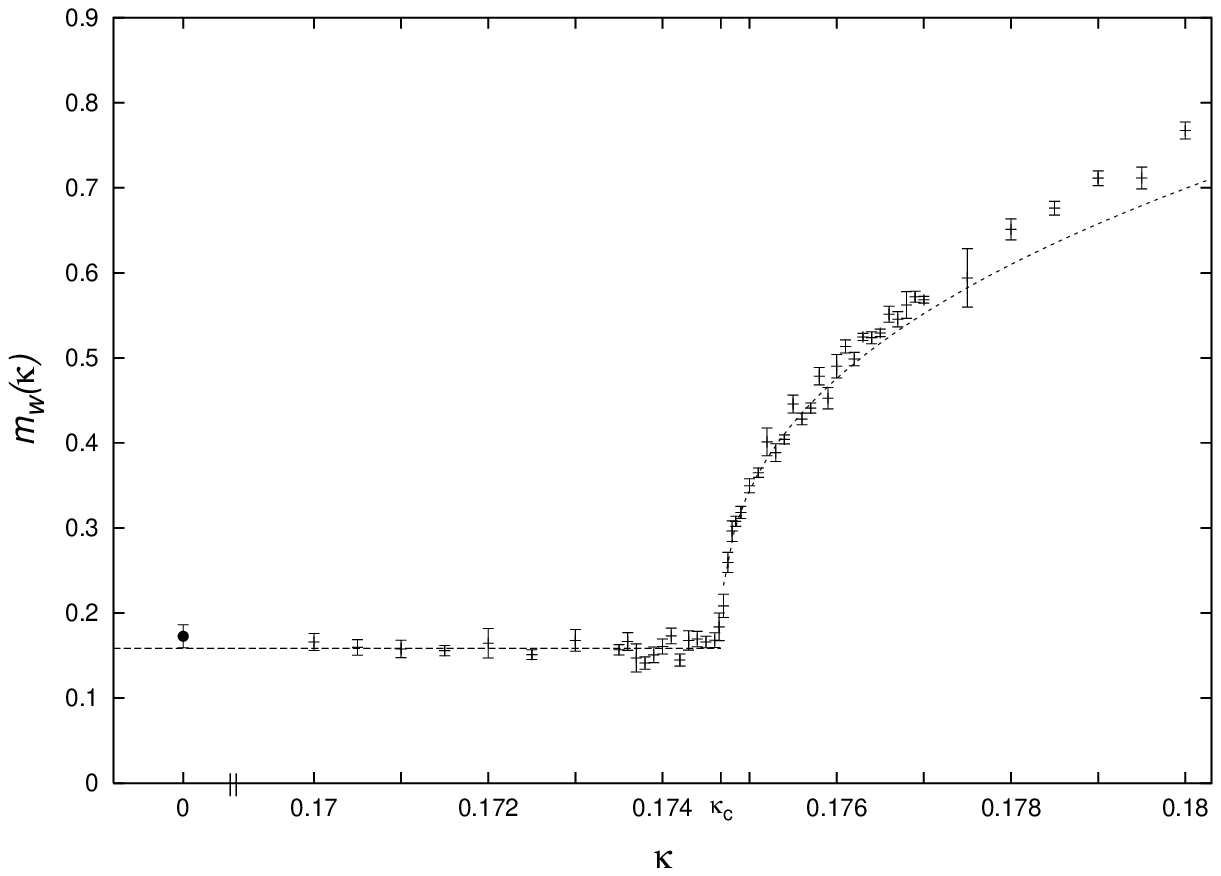, height=80mm}
\end{center}
\caption{Propagator masses obtained from
fits to correlation functions calculated on lattices of size $16^2\times 32$.
The curves show fits as explained in the text. The filled circle at
$\kappa=0$ gives the result for the pure $SU(2)$ gauge theory. }
\label{propmasses.fig}
\end{figure}

From the above analysis of finite size effects on propagator masses below
and above $\kappa_c$ we conclude that masses can reliably be extracted from
correlation functions already on lattices of size $16^2 \times 32$ using
a fit of the form given in Eq.~(\ref{wfit}). The results obtained this way for
a large number of $\kappa$ values are shown in Fig.~\ref{propmasses.fig}. We
note that within our numerical accuracy the propagator mass $m_w$ is
independent of $\kappa$ in the symmetric phase while it rises rapidly above
$\kappa_c$. We also have performed a calculation at $\kappa =0$, $\lambda_3 =
0$, i.e. in the pure $SU(2)$ gauge theory. This yields a value for the
propagator mass which is consistent with those obtained in the symmetric
phase of the $SU(2)$ gauge-Higgs model close to $\kappa_c$. This pure gauge
value also is shown in Fig.~\ref{propmasses.fig} as a filled circle.
A fit to the data for $m_w$ below $\kappa_c$ yields

\beqn
m_w = 0.158 \pm 0.002 \quad, \quad \kappa \le \kappa_c~.
\label{mwfitb}
\eqn
Above $\kappa_c$ the mass increases rapidly. A good fit to the data in the
entire range $\kappa \ge \kappa_c$ is obtained with the ansatz

\beqn
m_w = 0.158 + a (\kappa - \kappa_c)^\beta \quad \kappa \ge \kappa_c~.
\label{mwfita}
\eqn
In order to be more sensitive to the critical behaviour close to $\kappa_c$
we have restricted the fit to the interval $\kappa_c \le \kappa \le 0.176$.
In this case we find for the two free parameters $a=4.0(4)$ 
and $\beta = 0.384(15)$.
We note that the exponent $\beta$ turns out to be consistent with that
of the $O(4)$ spin model in 3-dimensions. In a recent
Monte Carlo analysis \cite{Kan95} this exponent has been found to be
$\beta = 0.3836(46)$ which is in agreement with
results obtained from the $(4-\epsilon)$-expansion.
Through the Higgs-mechanism the $W$-boson mass in the $SU(2)$
gauge-Higgs model is linked to the scalar field expectation value.
It thus seems plausible that also the temperature dependence of the
$W$-boson mass close to $\kappa_c$ is controlled
by the exponent $\beta$. Although the agreement is quite striking we
stress that without a more detailed finite size analysis we can, at
present, not rule out a smooth crossover as found in Ref.~\cite{Buc95}
or a critical behaviour controlled by a one-component scalar field as
suggested in Ref.~\cite{Kaj95}.

The results for the propagator mass discussed so far are in good
quantitative agreement with analytic calculations of Ref. \cite{Buc95}.
Expressing our result in terms of $g^2$
we find in the symmetric phase $m_w = 0.35(1) g^2T$. This should be compared
with the approximate value $m_w \simeq 0.28g_3^2$ quoted in Ref.~\cite{Buc95} for
large values of $\lambda /g^2$. Also the functional form of $m_w (\kappa)$
in the symmetry broken phase is in good agreement with results obtained from
the analysis of gap equations when the $\kappa$-dependence is transformed
into temperature dependence with help of Eq.~(\ref{kappa}).
However, the emerging physical picture is rather different from what is
proposed in \cite{Buc95}.

Based on an analysis of coupled gap equations for the scalar
and vector propagators in Landau gauge it has been concluded in \cite{Buc95}
that also above
$T_c$ the propagator mass is determined by the vacuum expectation value of
the Higgs field. This expectation value is actually much smaller than at $T=0$
(mini-Higgs mechanism) and is proportional to $g^2$. For
scalar and gauge couplings corresponding to equal Higgs and $W$-boson masses
at zero temperature they do find a smooth crossover between the low and high
temperature regimes which are distinguished by different magnitudes of
the scalar field expectation values. However, it also is found that the high
temperature  magnetic mass is rather insensitive to the actual value
of $\lambda /g^2$.

Our calculation suggests a different physical mechanism. Since the pure gauge
propagator mass is compatible with the result of the Higgs-model in the
symmetric phase, we believe
that the magnetic mass is of {\it fully thermal origin}, without any high T
(low $\kappa$) Higgs effect. Above $\kappa_c$ a non-analytic contribution
adds to the thermal value which can be attributed to the onset
of the Higgs-effect in the low temperature phase. Although we can, at
present, not rule out a smooth crossover behaviour we note that the
temperature dependence in the vicinity of $\kappa_c$ is well described
by a non-analyticity characteristic for continuous phase
transitions in Higgs models.

%
\section{Gauge invariant vector and scalar correlators}

In the symmetry broken phase the $W$-boson propagator mass increases
rapidly with $\kappa$.  This tendency
agrees well with both the low temperature vector mass extracted from
the gap equation (when $g \simeq 2/3$
and $a=1/T$ of our simulation are taken into account) and with the masses
obtained from gauge invariant correlators in lattice units by
\cite{Ilg95,Kaj95,Phi96}. These latter simulations do, however, use a
smaller value for $\lambda$  than we do. It therefore is important to
compare for
our choice of couplings the agreement of the masses extracted from gauge
invariant vector operators with the propagator mass.

We have calculated correlators for the standard gauge
invariant vector and scalar operators
\beqn
O_{v,i} (x) = {\rm Tr} \sigma_3 \Phi^\dagger_xU_{x,i}  \Phi_{x+\hat{i}}
\quad i=1,~2\quad ,
\label{cordefv}
\eqn
as well as two different scalar operators,
\beqn
\quad O_s^a =  {\rm det} \Phi_x  ~~,
\quad O_s^b =  \sum_{i=1}^2{\rm Tr} \Phi^\dagger_x U_{x,i} \Phi_{x+\hat{i}}  ~~,
\label{cordefs}
\eqn
In analogy to Eq.~(\ref{corr}) we define operators $\tilde{O}$ which project
onto zero momentum states.
The long distance behaviour of the scalar correlator $G_s^\alpha (z) =
\langle \tilde{O}_s^\alpha (0) \tilde{O}_s^\alpha (z) \rangle$, 
$\alpha=a,~b$, does then yield the scalar (Higgs) mass while the correlator
$G_v (z) = \sum_{i=1,2}\langle \tilde{O}_{v,i} (0) \tilde{O}_{v,i} (z) \rangle $
defines the mass of a vector particle with the quantum numbers of the $W$-boson.

\begin{figure}[htb]
\begin{center}
  \epsfig{file=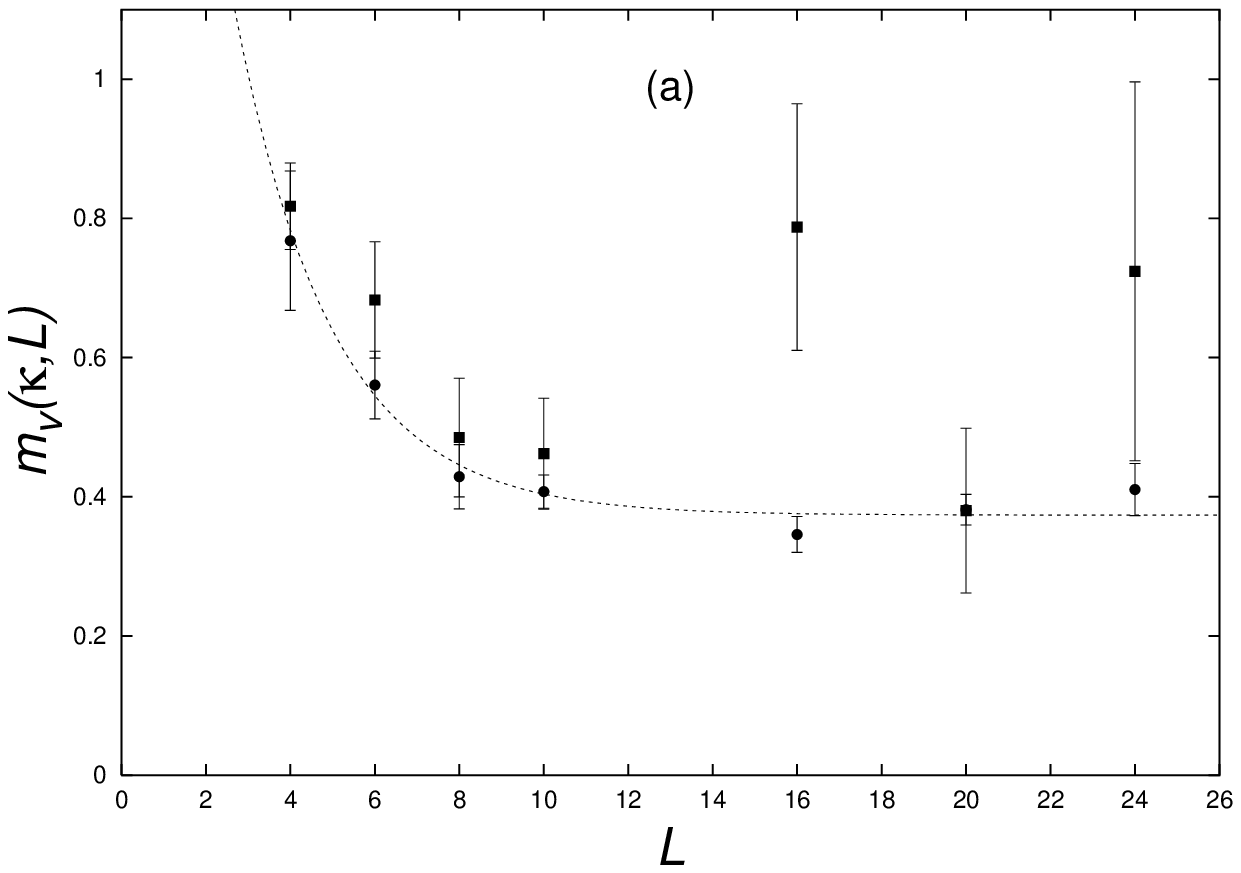, width=73mm, height=65mm}
  \hfill
  \epsfig{file=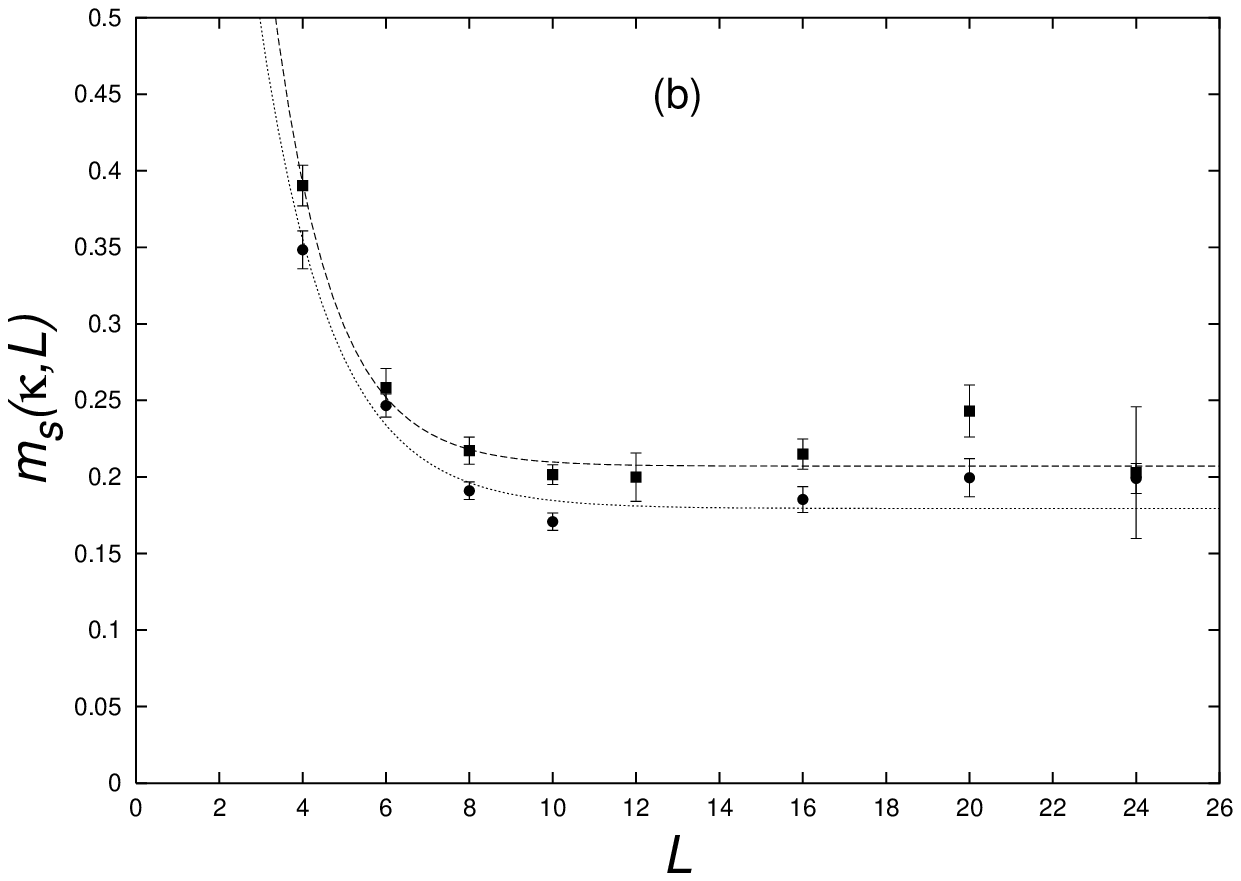, width=73mm, height=65mm}
\end{center}
\caption{Vector ($a$) and scalar ($b$) masses at $\kappa = 0.1745$ (squares)
and 0.17484 (circles) obtained
from fits to gauge invariant correlation functions calculated on lattices
of size $L^2\times 32$ with $L$ ranging from 4 to 24. Curves show
exponential fits for the volume dependence of the masses. }
\label{localmsw.fig}
\end{figure}

In both cases we also have performed an analysis of the finite size
dependence of masses extracted from these correlation functions. Some
results are given in Fig.~\ref{localmsw.fig}. This shows that also in this
case the masses may reliably be analyzed on lattices of size $16^2 \times
32$.

\begin{figure}[htb]
\begin{center}
  \epsfig{file=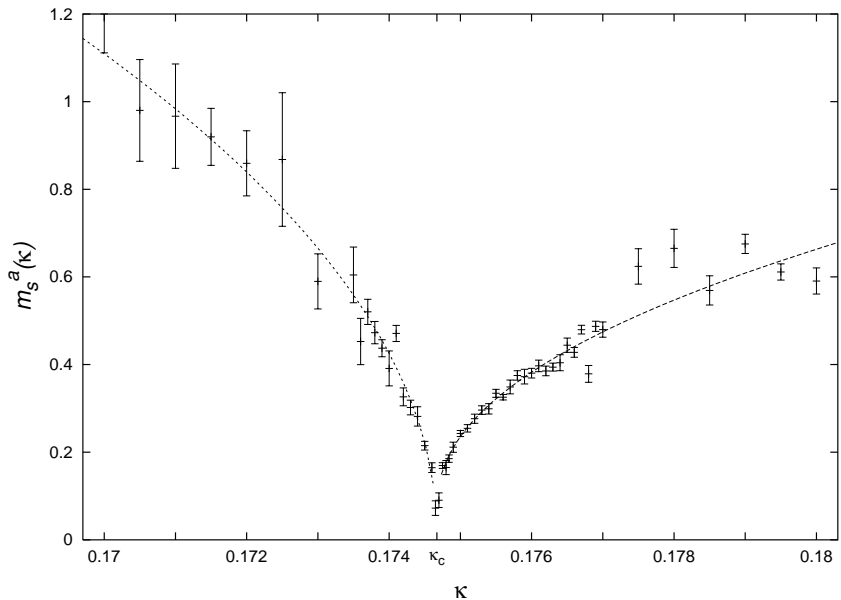, height=80mm}
\end{center}
\caption{Scalar masses obtained from fits to the gauge invariant correlation
function $G_s^a$ calculated on lattices of size $16^2\times 32$. }
\label{scalarmass.fig}
\end{figure}

The scalar (Higgs) mass has been calculated on the same set of
configurations used for the analysis of the $W$-boson propagator in Landau
gauge. Fits have been performed for distances $z\gsim (2-4)$.
We do find quite a different behaviour for the scalar mass in the
symmetric phase while it is similar in magnitude and functional dependence
to the propagator mass in the symmetry broken phase. In all cases we
obtained consistent results from the two scalar operators defined in
Eq.~(\ref{cordefs}). The masses extracted from $G_s^a$ are shown in
Fig.~\ref{scalarmass.fig}.

The scalar mass becomes very small at $\kappa_c$. The behaviour is consistent
with a second order or very weak first order phase transition. The
functional dependence of the scalar mass below and above $\kappa_c$ is
clearly different. We therefore have fitted the masses to the ansatz
\beqn
m_s = c_{\pm} + a_{\pm} |(\kappa - \kappa_c)|^{\nu_{\pm}} \quad,
\label{mwfits}
\eqn
where the subscript +/- refers to the broken/symmetric phases. We find
consistent fits for $c_{\pm} \equiv 0$ as well as $c_{\pm} \ne 0$. The fit
parameters for both cases are summarized in Table~\ref{fits.tab}.
\begin{table}
\begin{center}
\begin{tabular}{|l|l|l|l|l|l|}
\hline
\multicolumn{3}{|c|}{$\kappa \le \kappa_c$} &
\multicolumn{3}{|c|}{$\kappa \ge \kappa_c$} \\
\hline
\multicolumn{6}{|c|}{fit with $c_\pm \ne 0$} \\
\hline
$c_-$ & $a_-$ & $\nu_-$ &$c_+$ & $a_+$ & $\nu_+$  \\
\hline
0.04~(2)  &  26~(9)  & 0.58~(6) & 0.073~(20)  & 6~(2)  & 0.45~(6) \\
\hline
\multicolumn{6}{|c|}{fit with $c_\pm \equiv 0$} \\
\hline
-  &  14~(2)  & 0.48~(4) & -  & 3.0~(5)  & 0.31~(3) \\
\hline
\end{tabular}
\end{center}
\caption{Results of the fits to the masses extracted from the scalar correlation
function $G_s^a$.
In Fig.~6.  we have shown the fits with $c_\pm \equiv 0$.}
\label{fits.tab}
\end{table}
We note that also this behaviour is quite similar to the results obtained from
gap equations \cite{Buc95}.

From Table~\ref{fits.tab} we conclude that the determination of critical
exponents is quite sensitive to the inclusion or exclusion of a constant
term in the fits. This shows that also here we will need a rather
detailed finite size analysis close to
$\kappa_c$ to draw definite conclusions. Still a few observations may be
appropriate already at this point: When approaching
$\kappa_c$ from below (in the symmetric phase) we find that $\nu$ is consistent
with the mean field
value 1/2. This has also been observed in another simulation of the 
3-dimensional model \cite{Kaj95}. 
However, when approaching $\kappa_c$ in the broken phase we
find a smaller value for the exponent $\nu$. This is particularly
true when we exclude the constant term in our fits. In any case, it does seem
that a large exponent, i.e. $\nu > 1/2$, like in the 3-d Ising model
($\nu \sim 2/3$) or the $O(4)$-model ($\nu \simeq 0.75$ \cite{Kan95}) is
ruled out by our data. It seems that the temperature dependence of the
scalar mass is very similar to that of the $W$-boson propagator mass,
although in the case of a second order phase transition both should
depend on different critical exponents. As we have argued in the
previous section the temperature dependence of the $W$-boson propagator
mass is expected to be controlled by the exponent
$\beta$ while the scalar mass is controlled by the correlation length
exponent $\nu$.
This may hint at are more complex dynamics close to $\kappa_c$ than
described by the universality class of 3-$d$, scalar spin models and may
even indicate the possible existence of a tricritical point. 

A similar detailed analysis of the gauge invariant vector correlation function
on our data sets failed because the signal disappeared already at
rather
short distances ($z\simeq 4$) in the statistical noise. The construction of
improved operators may help in this channel \cite{Phi96}.
We could calculate the vector mass at the two $\kappa$ values close
to $\kappa_c$, where we did perform the finite size analysis (see
Fig.~\ref{localmsw.fig}). Here we extracted
the masses in the vector channel from a fit to the correlation functions for
$z \ge 2$. This yields
\begin{eqnarray}
m_v &=& \cases{ 0.557 \pm 0.087 \quad,\quad \kappa = 0.1745 \cr
0.356 \pm 0.028 \quad,\quad \kappa = 0.17484 }
\label{vectormass}
\end{eqnarray}
We note that in the symmetric phase at $\kappa=0.1745$ the mass in the
vector channel is more than twice as large as the mass extracted from
the $W$-boson propagator ($m_w \simeq 0.158$), whereas these masses are
similar in the symmetry broken phase. For instance, we find 
from Table~\ref{corrfit.tab} at $\kappa = 0.17484$ for the $W$-boson
propagator mass $m_w = 0.308~(6)$, which is
compatible with the value given in Eq.~\ref{vectormass} for $m_v$.

\section{Conclusions}
In this paper we have established the existence of the exponential
decay of gauge field (link-link) correlations of the 3-d gauge-Higgs system 
in Landau gauge which leads to a non-vanishing magnetic screening mass in 
the $W$-boson propagator. In the high temperature (small $\kappa$) regime
this characteristic length scale agrees with the same quantity of
the pure gauge system.

This equality cannot be a coincidence, and gets further support from
the study of the gauge invariant excitation spectrum by Philipsen {\it et al.}
\cite{Phi96}. They report the non-mixing of a $0^{++}$ state composed of
gauge plaquettes with those operators having the same quantum numbers
and also involving Higgs fields. This decoupling phenomenon is actually
expected at high temperature; it corresponds to the separation of the heavy
scalar modes from the dynamics of the weakly screened magnetic fluctuations 
which are described by an  
effective theory both in the case of QCD and the gauge-Higgs system
\cite{Bra95}.

On the basis of this apparent decoupling we have argued that the magnetic 
vector fluctuations do not receive any contribution to their screening mass
from a Higgs-type mechanism in the high temperature phase. 
The onset of the additional mass generation through the Higgs-mechanism can be
observed as a well-localized increase of the effective mass above $\kappa_c$.
Our present calculations, which have been performed with a set of couplings
corresponding to $m_H\approx 80GeV$, suggest the existence of a second order
phase transition. While the temperature dependence of the $W$-boson
propagator mass close to $\kappa_c$ is consistent with the critical behaviour 
expected from an $O(4)$ symmetric effective theory, we seem to find
deviations from this picture for the scalar mass.
However, only a very careful finite size analysis can substantiate this
observation and should allow to 
distinguish from a smooth crossover suggested by \cite{Buc95} or the
critical behaviour of an effective theory possibly controlled by a 
one-component scalar field as suggested in \cite{Kaj95}.

The propagator masses and the gauge invariant spectrum agree well in the
broken symmetry phase.

An important issue is to clarify why the two kinds of operators which yield the 
same mass in the symmetry broken phase cease to couple to the same state in the
symmetric phase. Further investigations of gauge invariant and gauge
dependent correlation functions should lead to progress on this
question. One possibility would be, for instance, to construct also simple 
non-gauge invariant two-particle operators whose correlators in the Landau 
gauge could reproduce the results of the gauge invariant spectroscopy.

Another important next step towards the clarification of the nature of the
symmetric phase and its fundamental degrees of freedom is the thorough 
investigation of the contribution
of the static sector to the equation of state of the finite temperature
gauge-Higgs system. Our analysis suggests that the thermodynamics 
in the symmetric phase may be described in terms of almost free massive
degrees of freedom having the mass explored in the present paper.

\medskip
\noindent
{\bf Acknowledgements:}
The computations have been performed on Connection Machines at the
H\"ochstleistungs\-rechenzentrum (HLRZ) in J\"ulich, the University of
Wuppertal and the Edinburgh Parallel Computing Center (EPCC).
We thank the staff of these institutes for their support.
The work of FK has been supported through the Deutsche
Forschungsgemeinschaft under grant Pe 340/3-3. JR has partly been supported
through the TRACS program at the EPCC. A.P. acknowledges a grant from OTKA.


\begin{thebibliography}{99}

\bibitem{Lin72} A.D. Linde and D.A. Kirzhnits, Phys. Lett. {\bf B72} (1972)
471.

\bibitem{Arn93} O. Espinosa and P. Arnold, Phys. Rev. {\bf D47} (1993)
3546.

\bibitem{Karun} F. Karsch and A. Patk\'os (unpublished).

\bibitem{Arn96} P. Arnold and L.G. Yaffe, Phys Rev {\bf D52} (1996) 7208.

\bibitem{Buc94} W. Buchm\"uller, Z. Fodor, T. Helbig and D. Walliser,
Ann. Phys. (N.Y.) {\bf 234} (1994) 260.

\bibitem{Esp93} J.R. Espinosa, M. Quir\'os and F. Zwirner, Phys. Lett.
{\bf B314} (1993) 206.

\bibitem{Lin80} A.D.~Linde, Phys. Lett. {\bf 96B} (1980) 289.

\bibitem{Kal92} O.K. Kalashnikov, Phys. Lett. B279 (1992) 367.

\bibitem{Bra95} E. Braaten, Phys. Rev. Lett. {\bf 74} (1995) 2164.

\bibitem{Fod95} Z. Fodor, J. Hein, K. Jansen, A. Jaster and I. Montvay,
Nucl. Phys. {\bf B439} (1995) 147.

\bibitem{Ilg95} M. Ilgenfritz, J. Kripfganz, H. Perlt and A. Schiller, Phys.
Lett. {\bf B356} (1995) 561.

\bibitem{Kaj95} K. Kajantie, M. Laine, K. Rummukainen and M. Shaposhnikov,
{\it The Electroweak Phase Transition: A Nonperturbative Analysis},
CERN-TH/95-263, October 1995.

\bibitem{Phi96} O. Philipsen, M. Teper and H. Wittig,
{\it On the Mass Spectrum of the SU(2) Higgs Model in 2+1 Dimensions},
Oxford-preprint, OUTP-95-40P, February 1996.

\bibitem{Buc95}
W. Buchm\"uller and O. Philipsen, Nucl. Phys. {\bf B443} (1995) 47.

\bibitem{PhiSint}
O. Philipsen,
in: Electroweak Physics and the Early Universe,
NATO ASI Series B:Physics Vol. 338, p. 393, (Edts. J.C. Romao and F.
Freire), Plenum Press 1994.

\bibitem{Gro94}
B. Grossmann, S. Gupta, U.M. Heller and F. Karsch,
Nucl. Phys. {\bf B417} (1994) 289.

\bibitem{Hel95}
U.M. Heller, F. Karsch and J. Rank, Phys. Lett. {\bf B355} (1995) 511.


\bibitem{Dos95} H.-G. Dosch, J. Kripfganz, A. Laser and M.G. Schmidt,
Phys. Lett. {\bf B365} (1995) 213.

\bibitem{Kaj96} K.Kajantie, M. Laine, K. Rummukainen and M. Shaposhnikov,
 Nucl. Phys. {\bf B458} (1996) 90.

\bibitem{Nak91}
A. Nakamura and M. Plewnia, Phys. Lett. B255 (1991) 274; \\
Ph. deForcrand, J.E. Hetrick, A. Nakamura an M. Plewnia,
Nucl. Phys. B (Proc. Suppl.) 20 (1991) 194; \\
E. Marinari, R. Ricci and C. Parrinello,
Nucl. Phys. B (Proc. Suppl.) 20 (1991) 199.


\bibitem{Kaj93} K. Kajantie, K. Rummukainen and M. Shaposhnikov, Nucl. Phys.
{\bf B407} (1993) 356.

\bibitem{Kaj94} K. Farakos, K. Kajantie, K. Rummukainen and M. Shaposhnikov,
Nucl. Phys. {\bf B425} (1994) 67.

\bibitem{Jak94} A. Jakov\'ac, K. Kajantie and A. Patk\'os, Phys. Rev. {\bf D49}
(1994) 6810.

\bibitem{Pol95} A. Patk\'os, P. Petreczky and J. Pol\'onyi,
                {\it Renormalisation Group Aided Reduction of Field Theories at
                Finite Temperature} (to appear in Ann. Phys. (N.Y.)),
                hep-ph/9505221.

\bibitem{Kar95}
F.~Karsch and J.~Rank,
\LAT {\bf 42} (1995) 508.

\bibitem{Dav88}
C.T.H.~Davies, G.G.~Batrouni, G.R.~Katz, A.S.~Kronfeld, G.P.~Lepage,
K.G.~Wilson, P.~Rossi, B.~Svetitsky,
\PR {\bf D37} (1988) 1581.

\bibitem{Man88}
J.E.~Mandula and M.~Ogilvie,
\PL {\bf B201} (1988) 117 and 
\PL {\bf B248} (1990) 156.

\bibitem{Nak95}
A. Nakamura, H. Aiso, M. Fukuda, T. Iwamiya, T. Nakamura and M. Yoshida,
{\it Gluon Propagators and Confinement}, hep-lat/9506024.


\bibitem{Kan95} K. Kanaya and S. Kaya, Phys. Rev. D51 (1995) 2404.

\end{thebibliography}
\end{document}